# *Superlattices based on van der Waals 2D materials*

*Yu Kyoung Ryu[1,*], Riccardo Frisenda[1], Andres Castellanos-Gomez[1,*]*

[1]*Materials Science Factory. Instituto de Ciencia de Materiales de Madrid (ICMM-CSIC), Madrid, E-28049, Spain.*

\* yukyoung.ryu@csic.es , andres.castellanos@csic.es

## Abstract

Two-dimensional (2D) materials exhibit a number of improved mechanical, optical, electronic properties compared to their bulk counterparts. The absence of dangling bonds in the cleaved surfaces of these materials allows combining different 2D materials into van der Waals heterostructures to fabricate p-n junctions, photodetectors, 2D-2D ohmic contacts that show unexpected performances. These intriguing results are regularly summarized in comprehensive reviews. A strategy to tailor their properties even further and to observe novel quantum phenomena consists in the fabrication of superlattices whose unit cell is formed either by two dissimilar 2D materials or by a 2D material subjected to a periodical perturbation, each component contributing with different characteristics. Furthermore, in a 2D materials-based superlattice, the interlayer interaction between the layers mediated by van der Waals forces constitutes a key parameter to tune the global properties of the superlattice. The above-mentioned factors reflect the potential to devise countless combinations of van der Waals 2D materials based superlattices. In the present feature article, we explain in detail the state-of-the-art of 2D materials-based superlattices and we describe the different methods to fabricate them, classified as vertical stacking, intercalation with atoms or molecules, moiré patterning, strain engineering and lithographic design. We also aim to highlight some of the specific applications for each type of superlattices.

# 1. Introduction

In their *Communication* (1970), IBM researchers Esaki and Tsu envisioned theoretically the realization of a novel semiconductor structure[1], which consisted either in a periodic variation of doping level in a single material or a periodic variation of two dissimilar materials. They claimed that if the period of such structures were shorter than the electron mean free path, it would be possible to observe new quantum confinement phenomena, due to the opening of extra allowed and forbidden energy bands, not available in a traditional device. They coined this novel device with a periodic potential as "*superlattice*". Shortly after, Esaki and Chang demonstrated quantum transport states in a 50-period GaAs-AlAs superlattice fabricated by molecular-beam epitaxy[2]. Later, Osbourn introduced the strain-engineering as a new variable to modulate the bandgap in a superlattice by controlling the growth of alternative layers with a certain lattice mismatch[3]. In these pioneering works, the authors had predicted that the experimental evolution on superlattices would mean the opening of new fields in semiconductor research. Since then, the semiconductor industry has developed standard fabrication methods to obtain superlattices from conventional 3D semiconducting materials through molecular beam epitaxy, sputtering or other vacuum deposition methods[4–6]. These conventional superlattices have reached a limitation in their thermoelectric, optoelectronic, energy-storage performance that does not meet the current needs of the society. On this basis, the nanowires, due to their reduced dimensions and high surface-to-volume ratio, are being considered as new building blocks for the next generation of superlattices[7–9]. Nevertheless, the control over the interface states to obtain an optimal, abrupt junction is difficult due to the covalent nature of most of the nanowire heterojunctions and the presence of dangling bonds in their surface[10].

The isolation of 2D materials by mechanical or chemical exfoliation of bulk van der Waals materials and the synthesis of ultrathin van der Waals layered materials have changed one step further the paradigm in the fabrication of superlattices. 2D materials are atomically thin and have an area in the order of hundreds microns square. This trait endows them extremely high surface-to-volume ratio. The surface of these materials is devoid of dangling bonds and their heterostructures are held by van der Waals forces. As a result of these features, 2D materials have shown extraordinary physical and chemical properties. This has motivated the synthesis of novel 2D nanomaterials such as ultrathin metal-organic framework layers[11], transition metal oxides and hydroxides[12] or covalent-organic framework layers[13,14], in order to obtain enhanced chemical reactivity and device performance in comparison to their 3D, covalent homologues. This wide variety of 2D nanomaterials and available methods to prepare them open the door to the emerging field of van der Waals superlattices. Starting with 2D materials as the building blocks, periodic potentials from structures with precise atomic resolution can be now easily achieved, even without the need of vacuum deposition techniques. The exploration of superlattices based on van der Waals 2D materials, both theoretically and experimentally, has barely started. However, there exists already a solid body of results that contributes to settle down the bases of the field. These results are the focus of the present feature article, divided into the main 2D based superlattice types shown in table 1:

1. Vertical stacking of dissimilar 2D layers, where the periodical potential is modulated principally by the chemical composition of the layers and/or the lattice mismatch existing between them.
2. Multi-layered 2D materials that are subjected to an intercalation process. In this type of superlattices, the periodical potential can be modulated by the chemical composition of the layers, the re-structuration of the functionalized layers after the intercalation of the hosting species and the charge transfer that takes place between the host flakes and the atoms or molecules.
3. Moiré structures by twisting of two 2D material-based layers. When two equivalent (e.g. graphene bilayer) or dissimilar (e.g. graphene/h-BN) adjacent layers are rotated under certain angles, the lattice mismatch and/or the atom rearrangement give rise to moiré-pattern superlattices with new lattice periodicities.
4. Strain-engineered layers. A periodic potential can be produced by subjecting a 2D material flake to periodic tensile/compressive strains or by introducing a periodic, controlled mismatch between the lattices of two dissimilar 2D materials.
5. Synthetic superlattices defined by lithography. In this approach, the periodic potential can be induced in a 2D material by placing it on a patterned topography or by patterning the material itself with a lithographic technique.

| Classification | Elements of the unit cell | Modulation by (dimension) | Applications (references) |
|---|---|---|---|
| Vertical stacking | 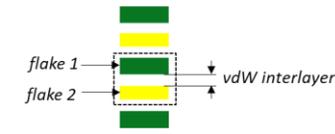 | Chemical composition, lattice (mis)match (1D) | • IR photodetectors (4,5,9)<br>• Phase-change memory (17 – 23)<br>• Thermoelectrics (36 – 39)<br>• Superconductivity (40 – 43) |
| Intercalated compounds | 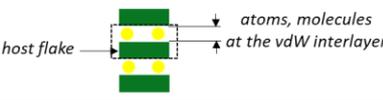 | Chemical composition, structuring, charge transfer (1D) | • Optoelectronics (44 - 50, 61 – 64)<br>• Superconductivity (51)<br>• Thermoelectrics (52 - 54)<br>• Energy storage (56 - 60) |
| Moiré periodic patterns | 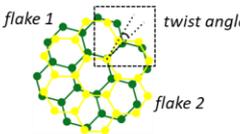 | Atoms rearrangement, misorientation angle, strain (2D) | • Diract point states (76, 84 - 88)<br>• Hofstadter's butterfly (78 - 83)<br>• Superconductivity (89 - 95)<br>• Moiré interlayer excitons (96 - 98) |
| Strain engineered | 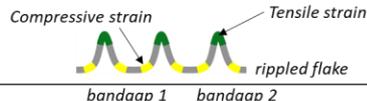 | Periodic in-plane or out-of-plane strain (1D or 2D) | • Phase-change memory (106, 107)<br>• Bandgap engineering (108 – 111)<br>• Optoelectronics (114, 115) |
| Lithographic engineered | 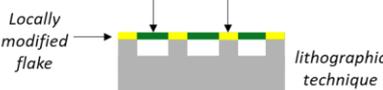 | Dielectric patterning, periodic substrate, Periodic patterning, strain (1D or 2D) | • Artificial crystals (117 – 119)<br>• Moiré periodicities (120)<br>• Bandgap engineering (121, 122) |

Table 1. Type of van der Waals 2D material superlattice (first column), scheme of the elements that compose their unit cell (second column), mechanisms to modulate the superlattice for each type (third column) and their applications found in the literature up to now (fourth column).

Note that in some cases there is more than one single mechanism involved in the fabrication of the superstructure. Specifically, either as an explicit parameter or produced unintentionally, strain is found to have an important role in the modulation of electronic/optical properties in most of the other types of superlattices.

Each type of superlattice (1 to 5) and its potential applications will be discussed in detail in the next section and the corresponding subsections.

## 2. 2D material superlattices

### 2.1. Vertically stacked superlattices
#### 2.1.1. Naturally occurring superlattices

Natural van der Waals superlattices have been obtained by isolating thin layers from minerals such as franckeite [15–19] or cylindrite [20]. Both minerals belong to a family of sulfosalts with generic formula $(Pb,Sn)^{2+}_{6+x}Sb^{3+}_2Fe^{2+}Sn^{4+}_2S_{14+x}$, with $-1 \leq x \leq 0.25$ [21]. Franckeite constitutes a natural misfit compound mineral, whose superlattice structure has been formed by alternating phase segregation of pseudo-hexagonal Sn-rich and pseudo-tetragonal Pb-rich layers, separated by a van der Waals gap (Figure 1(a)). The electrical characterization of exfoliated flakes from this material shows that is a highly p-type doped semiconductor with a narrow bandgap ($< 0.7$ eV), which is suitable for infrared (IR) detection [15,16]. In the case of cylindrite, thin flakes were obtained from mechanical or liquid phase exfoliation. The characterization by high-resolution TEM images showed that the cylindrite thin flakes were composed of a stack with alternating pseudo-tetragonal Pb-rich and octahedral Sn-rich layers. Analogous to the franckeite thin layers, the thin cylindrite layer-based field-effect transistors behaved as a highly p-type doped semiconductor with a narrow bandgap (<0.85 eV), suitable to measure photoresponse in the IR. The characterization of the magnetic properties of the thin flakes proved that the intrinsic magnetic interactions present in the bulk mineral are preserved after thinning [20]. Both minerals have the advantage to be stable under ambient conditions. These results underline the existence of more natural minerals from which thin layers with novel properties can be isolated.

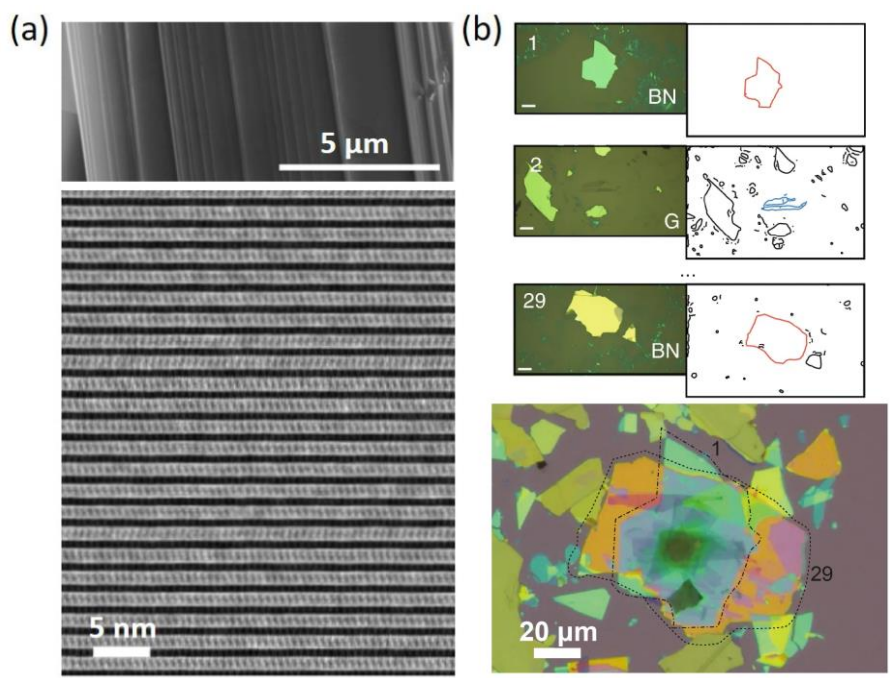

Figure 1 (a) SEM image of an exfoliated franckeite showing the layered nature of the mineral (top). Cross-section TEM image of the superlattice viewed along the [100] direction. Reproduced with permission of ref. 5. (b) Artificial van der Waals superlattice comprising 29 alternating layers of graphite and h-BN. Top: optical microscope images of some of the graphite (G) and BN flakes utilized to assemble the superlattice (left) and the vector-drawing line-edges of the flakes automatically extracted from the optical microscope images. The scale bars correspond to 20 μm. Bottom: optical image of the G/BN superlattice. The dashed lines highlight the bottom and top BN flakes. Reproduced with permission of ref. 12.

### 2.1.2 Artificial stacking

The fabrication of van der Waals 2D material superlattices by artificial stacking presents the advantage of choosing the materials and piling them up with an optimized interlayer interface to achieve specific optoelectronic properties. Kang *et al* developed a large-scale layer-by-layer vertical assembly process[22]. The authors achieved a large-area superlattice composed of nine alternating layers of $MoS_2$ and $WS_2$ prepared by sequentially peeling with thermal release tape and stacking the layers under vacuum. They showed the ease of detachment of this superlattice by peeling it off from the substrate in the water. In order to pave the way to manufacture complex superlattices with good quality and minimum time and human operation cost, an autonomous robotic assembly system was developed recently by Masubuchi, Machida and co-workers [23]. The localization, shape and thickness of exfoliated 2D materials are factors that must be taken into account to select and fabricate useful heterostructure devices. The authors estimated the time needed to fabricate manually a 29-layered superlattice in about 120 h of uninterrupted processing. Their system consisting in an automatic scanning of the flakes by optical microscope, recording of the position and geometry of the flakes by a computer-assisted design algorithm database and robotic stamp assembling achieved the alternating 29-layered graphene/h-BN superlattice in less than 32 h, with a human operation time less than 6 h (Figure 1 (b)). Both works address the challenges of large-scale production and development of cost-effective manufacturing processes for moving the 2D-materials based devices to the industry level.

### 2.1.3 Van der Waals epitaxially grown superlattices

Molecular beam epitaxy is a high-vacuum, low deposition rate technique that is employed to grow semiconductor heterostructures, which present atomically smooth and abrupt interfaces. Due to these reasons, it is also being used to fabricate high-quality 2D materials-based superlattices where the different epi-layers are upheld by van der Waals forces[24]. Moreover, since the 2D cristalline layers do not present dangling bonds, each epi-layer grows without having to accommodate its lattice constant to that of the layer below. Therefore, some authors define the MBE on 2D materials as van der Waals epitaxy [25,26]. Hereafter, we detail the results achieved by both methods. Epitaxial growth methods allow the precise designing of the transport properties as a function of the chemical composition and thickness of the layers that constitute the superlattice. Recently, by reducing the thickness of the $Bi_2Se_3$ topological insulating layer in a $Bi_2Se_3$-$In_2Se_3$ superlattice from $(Bi_2Se_3)_{12}/(In_2Se_3)_6$ to $(Bi_2Se_3)_6/(In_2Se_3)_6$, the transport dimensionality was tunned from coherent 3D to incoherent 2D transport, respectively [27].

Interfacial phase-change memory materials that encompass the family of superlattices with the generic formula $[(GeTe)_x/(Sb_2Te_3)_y]_n$ are being studied to achieve fast switching and low energy consumption for non-volatile memory applications [28]. The TEM characterization of GeTe/$Sb_2Te_3$ superlattices fabricated by MBE and annealed at 400ºC revealed a reorganization of the superlattice structure into $Sb_2Te_3$ and GeSbTe rhombohedral layers [29]. The prediction of the authors about modulating the electrical, thermal or magnetic properties by controlling the MBE process in this kind of materials was accomplished by Cecchi *et al* [30]. The authors fabricated by molecular beam epitaxy $Sb_2Te_3$/$Ge_xSb_2Te_{3+x}$ superlattices. They observed that this configuration reduced the intermixing of the layers compared to the standard GeTe/$Sb_2Te_3$ superlattices fabricated in the previous work [29]. As a consequence, superlattices with less defective, sharper interfaces and higher carrier mobility were achieved. Another epitaxial method, radio-frequency sputtering, has been employed to fabricate the interfacial phase-change memory superlattices in several works [31–34]. Particularly, the phase change from a Dirac semimetal to a topological insulator as a function of the thickness of the GeTe layer has been observed, which is promising for spintronic applications [34].

In the case of natural minerals such as franckeite[15,16] and cylindrite[20], it was shown how the alternating formation of layers with dissimilar lattice symmetry gives rise to superlattices that grow with a mismatch, but are ordered and thermodynamically stable. This type of materials, called misfit layer compounds, can be synthesized with controlled properties as well. Their generic formula is $[(MX)_{1+\delta}]_m(TX_2)_n$, where M = Sn, Pb, Bi, Sb, rare earths; T = Ti, V, Nb, Ta, Cr and X = S or Se [35,36]. There is a variant of misfit layer compounds that is metastable and exhibits a turbostratic or rotational disorder along the c-axis, called ferecrystals. These materials present the generic formula: $[(MX)_{1+\delta}]_m(TX_2)_n$ where M = Sn, Pb, Bi, La; T = Ti, V, Nb, Ta, Cr, Mo, W and X = S, Se or Te [35–38]. In both formulas, the coefficients *m* and *n* represent the number of layers of each material per unit cell. In a misfit layer compound, there is only one axis of the layers that presents an incommensurate crystallographic direction (Figure 2(a), left. From [36]), while both in-plane axes of the layers in a ferecrystal are incommensurate and independent each other (Figure 2(a), right. From [36]). The dimensionality (*m* layers of material 1 and *n* layers of material 2 per unit cell, i.e. their ratio m/n) of these materials is taken into account to control the properties of both materials: the electrical transport [39–41], the charge transfer between the layers [42,43] or the presence/strength of a charge density wave state [44]. Another strategy followed to modulate the electrical properties is by substitution or doping in specific sites of the layers [45–47].

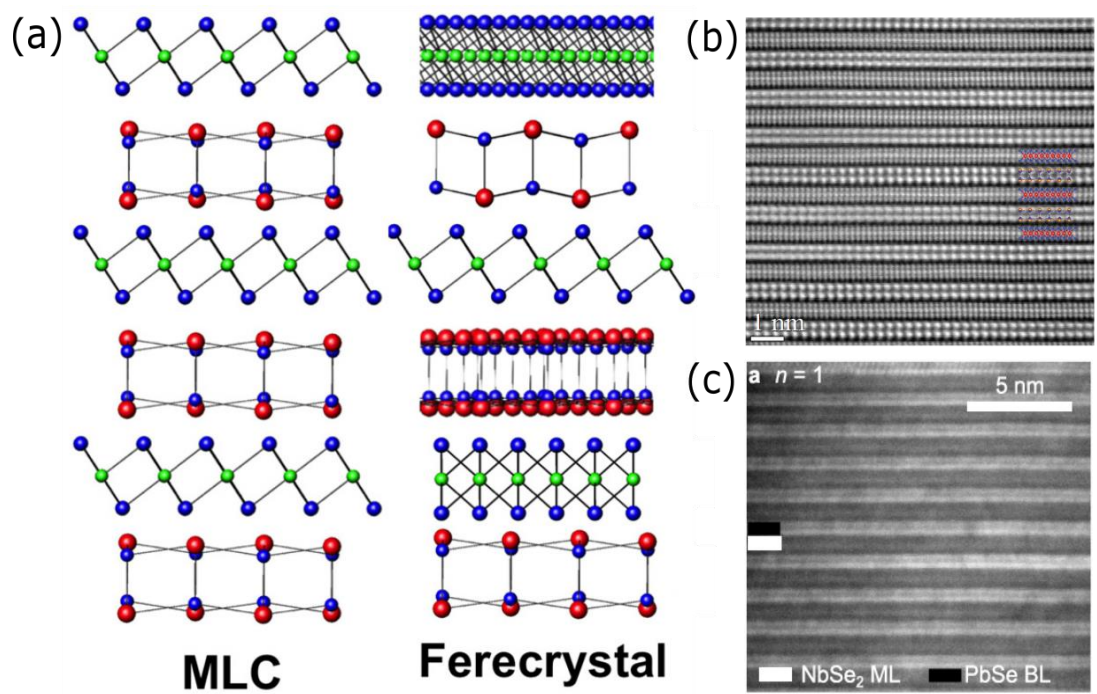

Figure 2 (a) Scheme of the superlattice structure of a misfit layerd compound (MLC, left) and a ferecrystal (right). Reproduced with permission of (ref. 25). (b) High resolution TEM image of a $(SnSe)_{1.16}(NbSe_2)$ MLC superconducting superlattice, viewed along the [010] direction. The inset is a scheme of the SnSe/NbSe$_2$ layers. Reproduced with the permission of (ref. 40). (c) Cross-section TEM image of a $(PbSe)_{1.14}(NbSe_2)_n$ ferecrystal superconducting superlattice with n = 1. Reproduced with the permission of (ref. 42).

Both misfit layer compounds and ferecrystals present a high potential in thermoelectric applications because their structure is ideal to develop phonon glass/ electron crystal responses, where dichalcogenide layers with high electrical mobility constitute the electron crystal component while the mismatch or turbostratic disorder along the c-axis lower the lattice thermal conductivity, acting as the phonon glass [36,48,49]. The recent work by Li *et al* [50] demonstrated the influence of the cross-plane thermoelectric properties as a function of the thickness of the layers that formed a misfit layer compound superlattice. The authors measured devices based on $(SnSe)_n(TiSe_2)_n$, varying the number of layers by *n* = 1, 3, 4, 5, but keeping the same global stoichiometry. They observed a quasi-linear increase in the cross-plane Seebeck coefficient from -2.5 to -31 µV/K as a function of *n*, and an approximately constant in-plane Seebeck coefficient for different values of *n*. Another strategy to improve the thermoelectric properties has been the Cu- and Co-substitution of Ti-sites in the TiS$_2$ layers of a $(SnS)_{1.2}(TiS_2)_2$ misfit-layered compound. An increased of the ZT value up to 33.3% in $(SnS)_{1.2}(Cu_{0.02}Ti_{0.98}S_2)_2$ was obtained compared to the pristine misfit layer compound. This improvement of the figure of merit was explained by terms of the increase in the effective mass and decrease in the carrier concentration induced by the Cu- incorporation and the decrease in the thermal conductivity induced by the disorder inherent to the misfit layer compound structuring [47].

Superconductivity is another application barely explored yet in both misfit layer compounds and ferecrystals. Superconductivity in the synthesized misfit layer compounds $(SnSe)_{1.16}(NbSe_2)$ [51] and $(SnS)_{1.15}(TaS_2)$ [52] with critical temperatures (T$_c$) of, respectively, 3.4 and 3.01 K, were recently reported. In the $(PbSe)_{1.14}(NbSe_2)_n$ ferecrystal [53], the shift of $T_c$ from 2.66 ($n = 3$), 1.91 ($n = 2$) and 1.11 ($n = 1$) K was observed. The authors correlated the reduction of the transition temperature with the dependence of the electron-phonon coupling and the interlayer charge transfer in the dimensionality (in this work, given by $n$) of the ferecrystal. The control of the transition temperature varying the $m$ index instead in a $[(SnSe)_{1+\delta}]_m(NbSe_2)$ ferecrystal has been also observed [54]. Figures 2(b) and 2(c) show respectively, the cross-section TEM images of a misfit layer compound [51] and ferecrystal [53] superconducting superlattice.

To summarize this section, the composition of the unit cell in terms of the number of layers and crystalline phase of each material determines the coupling and charge transfer between layers, which are the factors that most influence in the thermoelectric and electronic properties of this type of superlattices.

### 2.2 2D material layer-intercalated compound superlattices

In the previous section, it was discussed how the interlayer interactions between the layers that are part of a vertical stacking have a big influence in the cross-plane properties of the superlattice. Therefore, the coupling between layers for each system was adjusted by controlling the thickness, the doping and the crystallographic structure of the constituents. This peculiar interlayer spacing characteristic of van der Waals layered materials is found to have a considerable impact on the optical and electrical properties of the devices fabricated by or involved in intercalation methods. This section will review the works where ions or molecules where intercalated between the layers of a 2D material forming a different kind of superlattice.

A simultaneous enhancement in both the optical transmission and the electrical conduction through the phase change underwent by a 2D material after intercalation has been demonstrated in copper-intercalated bismuth-based chalcogenide layers [55,56]. The effect of Li intercalation on the charge density in different van der Waals heterostructures has been recently studied [57]. The tuning of the electrical and optical properties through intercalation has been also observed in ammonia ions-$WS_2$ host [58], lithium ions-$NbSe_2$ host [59], cetyl-trimethylammonium bromide (CTAB) in black phosphorus (BP) [60], and tetraethylbenzidine (EtDAB) ions in $PbI_2$ [61]. In a recent work, the coexistence of superconductivity and ferromagnetism originated from non-superconducting and non-ferromagnetic components has been achieved by the fabrication of a two-dimensional $SnSe_2$/ $Co(Cp)_2$ superlattice, where Cp are cyclopentadienyls molecules. A TEM cross-section of the superlattice is shown in figure 3(a) (left). The confinement effect that the $Co(Cp)_2$ molecules experience when they are sandwiched between the $SnSe_2$ layers weakens their coordination field, inducing the high spin ferromagnetic state (scheme in figure 3(a), right). The strong interlayer coupling between the Cp molecules and the $SnSe_2$ lattice induces a superconducting state through an unusual electron transfer [62].

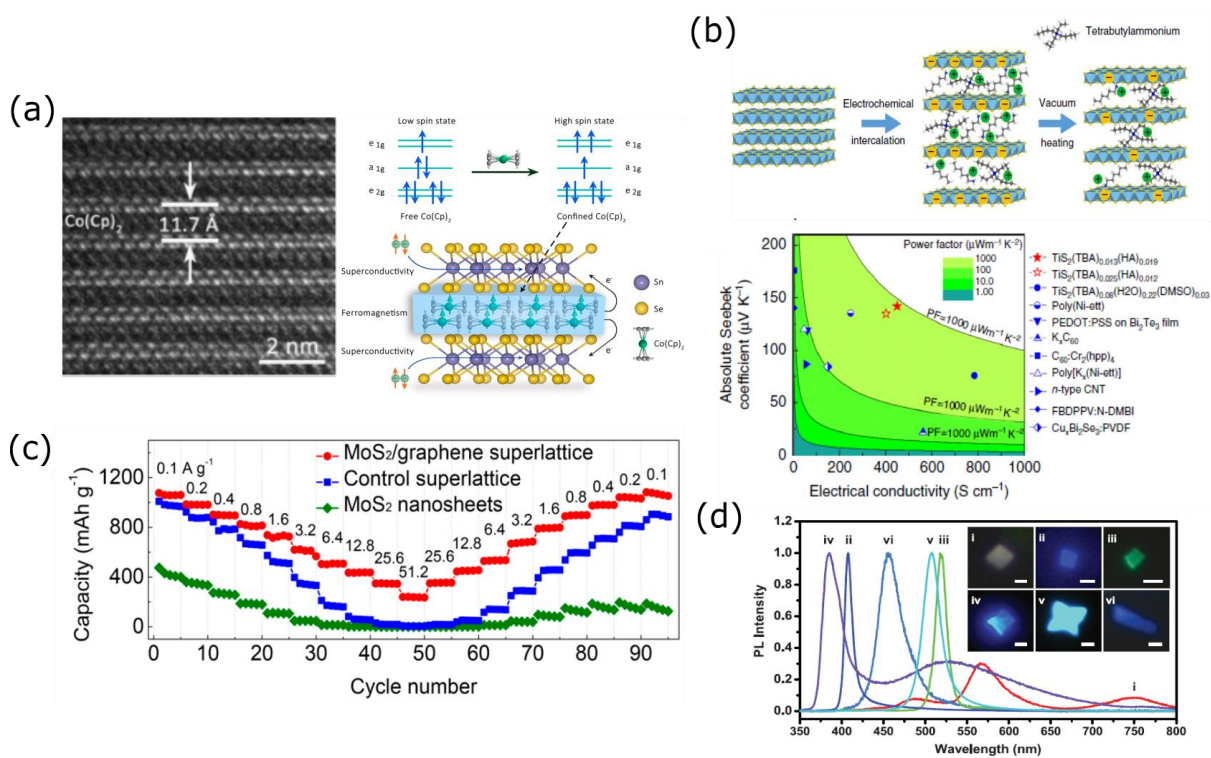

Figure 3 (a) Cross-section TEM image of the $SnSe_2$-$Co(Cp)_2$ superlattice (left). Scheme where the unit cell of the superlattice is shown, as well as the induced high spin state in the $Co(Cp)_2$ molecules when they are confined between the $SnSe_2$ layers (right). Reproduced with permission of ref. 51. (b) Scheme of one of the fabricated superlattices after electrochemical intercalation of hexylammonium/tetrabutylammonium molecules and vacuum heating (top). The absolute Seebeck coefficient represented as a function of the electrical conductivity shows that the intercalated $TiS_2$ compounds have a higher power factor compared to other thermoelectric materials (bottom). Represented with permission of ref. 52. (c) Rate capability as a function of cycling number of the 1T $MoS_2$/graphene superlattice is higher compared to control devices (1H $MoS_2$/graphene superlattice) and $MoS_2$ nanosheets. Reproduced with permission of ref. 57. (d) PL spectra of different 2D hybrid perovskites (corresponding to the crystals shown in the optical PL images of the inset). Optimizing the conditions of the same synthesis process, it is possible to achieve crystals with different wavelengths in the visible. Reproduced with permission of ref. 64.

In the previous section, we discussed how the interlayer disorder found in misfit layer compounds and ferecrystals could be controlled to optimize the carrier mobility and reduce the thermal conductivity to achieve high ZT for thermoelectric applications. Wan *et al* [63–65] employed another approach to achieve a similar effect in $TiS_2$. The authors intercalated at the interface of $TiS_2$ thin layers organic cations such as hexylammonium ions (HA) and polar organic molecules such as dimethyl sulfoxide (DMSO). A scheme of the fabrication process is shown in figure 3(b), top) [65]. The reason behind was to engineer the dielectric constant between the two-dimensional $TiS_2$ layers to optimize the electrical mobility/thermal conductivity ratio. With this hybrid inorganic-organic superlattice structure, the authors obtained a thermal conductivity 7 times lower compared to that of $TiS_2$ single crystal [64]. The further improvement of this process lead to the attainment of a power factor of 904 $\mu W/m^{-1}K^{-2}$, considerably higher than other thermoelectric materials, as showed in figure 3(b) (bottom) [65].

There is a research effort devoted to optimize 2D materials superlattices for energy storage applications [66]. In a smart design, $MoS_2$ monolayers were self-assembled by means of dopamine molecules adsorbed on their surface. After annealing, the dopamine layers sandwiched between the $MoS_2$ monolayers were thermally converted into nitrogen-doped graphene. This fabrication process results in a superlattice-like structure with good anode performance for lithium ion batteries. The fast charge-discharge rate and high cycling stability is attributed to the increment of reactive sites inherent of the new structure and an intimate interfacial coupling that enhances the electric and charge transfer [67]. In Sasaki's group, a superlattice based on metallic 1T-$MoS_2$/graphene monolayers has been fabricated by Li intercalation and solution-phase direct assembly. This superlattice was used as an anode for sodium ion batteries. It showed a larger capacity performance compared to the $MoS_2$ nanosheets and the 1H-$MoS_2$ monolayer/graphene monolayer superlattice (figure 3(c)) and a long-term stability up to 1000 cycles [68]. The same group has demonstrated the superior performance of metal oxide based anodes for lithium or sodium ion batteries when they are integrated in a van der Waals stacking superlattice. Two-dimensional metal oxides and hydroxides present optical properties and rich redox states that make them promising for energy storage applications. The authors fabricated a superlattice based on intercalated $MnO_2$ monolayers/graphene monolayers. The hybridization of the $MnO_2$/graphene monolayers resulted in an enhanced conductivity, achieving faster charge transport, a larger specific capacity and long-term stability up to 5000 cycles [69]. In a third approach, the same authors fabricated a $Ti_{0.87}O_2$ monolayers/graphene monolayers superlattice as an anode for sodium ion batteries, where the titanium oxide layer was synthesized with Ti vacancies to increase the number of redox active sites and the graphene layer to enhanced the conductivity [70]. When layered double hydroxide layers are inserted between $MoS_2$ monolayers, the strong interaction that takes place at the interface strengthens their electronic coupling, and thus, improved the electrical and charge transfer. As a result, the superlattice acts as a catalyst for overall water splitting [71].

The achievement of improved performance or novel phenomena from a given two-dimensional material by tunning its dimensionality and/or intercalating organic molecules in the interlayer space is also followed by the hybrid organic-inorganic perovskite field. Efficient blue LEDs at room temperature were fabricated from two-dimensional $(RNH_3)_2[CH_3NH_3PbX_3]_nPbX$ perovskites by (a) controlling the stacking number that gives rise to blue shift by quantum confinement and (b) designing dielectric quantum wells with organic host compounds to enhance the exciton binding energy and boost radiative recombination [72]. The control of the layer thickness in Ruddlesden-Popper type of perovskites, with the general formula of $(RNH_3)_2(CH_3NH_3)_{n-1}M_nX_{3n+1}$, has been studied to change the bandgap for achieving photodetectors with tunable wavelength [73–75] and to tune the cross-plane acoustic phonons transport [76]. In figure 3(d) it is shown the tunability of the wavelength as a function of the composition of 2D perovskites fabricated with the same synthesis process [75]. The level of anisotropy as a function of the $n$ inorganic layers in a 2D $EA_2MA_{n-1}Pb I_{3n+1}$ perovskite, where EA=$HOC_2H_4NH^{3+}$ and MA=$CH_3NH^{3+}$, was studied. The two dimensional perovskite with $n = 1$ and layer-edge 0º device displayed a humidity sensitivity almost four orders of magnitude higher than the 3D homologous perovskite [77].

In this section, ions and molecules intercalated between the layers of the host 2D material were the elements employed to modulate the interlayer coupling and charge transfer to obtain superlattices with improved electronic, optical and thermoelectrical properties and to enhance the chemical reactivity, fundamental for energy-storage applications.

## 2.3 Moiré superlattices

The main approach to obtain moiré superlattices is the stacking of two 2D layered materials with a controlled rotation angle or lattice mismatch. These two materials can be equal, such in the case of a twisted double-layer graphene [78] or the creation of a moiré superlattice from the rotation of the topmost layer of a $Bi_2Te_3$ quintuple layer [79]. Nevertheless, another common moiré superlattice is created by the stacking and twisting of two different materials, such as graphene on hexagonal boron nitride (h-BN) [80] or the van der Waals epitaxial growth of one of the materials over the other, such as $Bi_2Se_3$ on h-BN [81], GaSe monolayer on $MoSe_2$ monolayer [82], $MoS_2$ monolayer on $WSe_2$ monolayer [83] and $SnS_2$ monolayers on $WSe_2$ monolayers [84]. Recently, it has demonstrated the creation of a third moiré superlattice with a larger potential period from the overlap of the bottom and top moiré superlattices in a h-BN/graphene/h-BN system [85].

The number of novel quantum phenomena observed experimentally in graphene/h-BN Moiré superlattices is impressive and it is increasing fast. We address the readers interested in more details about this topic to a recent review [86]. The periodic potential created by the lattice mismatch between the graphene and the h-BN gives rise to new Dirac points which energy depends on the wavelength of the Moiré potential [87]. Two dimensional electrons under the simultaneous application of a magnetic field and a periodic electrostatic potential display a fractal-like quantized energy spectrum called Hofstadter's butterfly [88]. In 2013, the Hofstadter's butterfly was observed experimentally on graphene/h-BN lattices with small angle mismatches in the range of 4º - 15º by three different groups [89–91]. A scheme of the device and the experimentally observed Hofstadter's butterfly spectra are shown in figure 4(a) [89]. Later, one of these groups proved the coexistence of fractional quantum Hall effect states with integer states associated with the Hofstadter's butterfly spectrum and the apparition of novel fractional Bloch quantum Hall effects states at high magnetic fields, which conforms the complete unit cell of the butterfly spectrum [92]. Another of these groups have studied further the novel Bloch states that lead quantum oscillations, associated to the Hofstadter's butterfly [93,94]. The commensurate stacking leading topological bands and the incommensurate stacking leading non-topological bands (due to the inhomogeneous potential landscape induced by the moiré pattern) can be achieved by controlling the twist angle between graphene and h-BN [95,96]. A gate-tunable Mott insulator was demonstrated by varying the bandgap of an h-BN/trilayer graphene/h-BN moiré superlattice with a vertical electric field [97]. Novel plasmonic modes arisen from the Dirac mini-bands found in the moiré superlattice were observed [98]. Electron tunnelling spectroscopy was used to measure the energy gaps formed at the first and second Dirac points of mono- and bilayer graphene/h-BN moiré superlattices as a function of twist angles and external electric and magnetic fields [99].

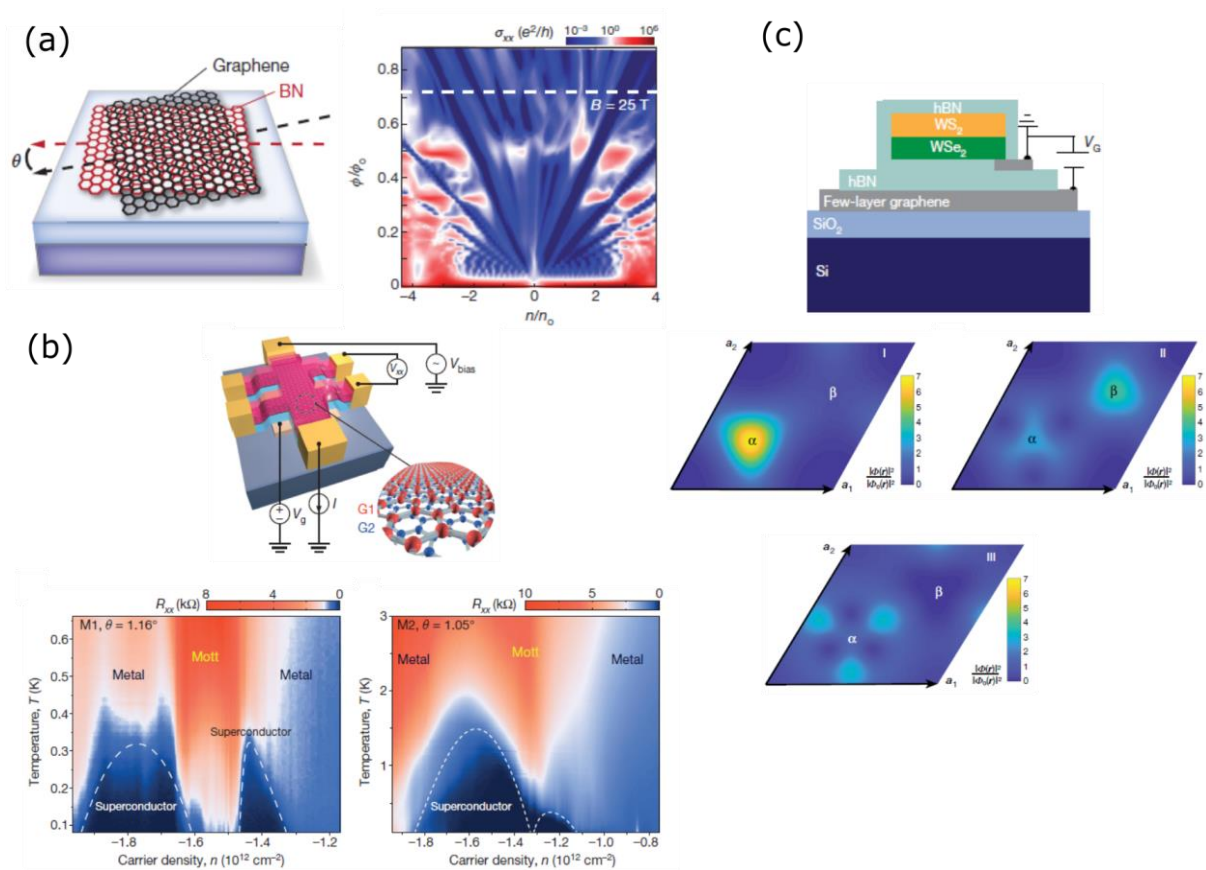

Figure 4. (a) Scheme of the twisted graphene and h-BN at an angle $\theta$. The moiré wavelength can be designed as a function of $\theta$ (left). Hall conductivity $\sigma_{xx}$ as a function of the normalized magnetic flux per unit cell and carrier density. A Hofstadter's butterfly with its characteristic fractal behaviour is displayed (right). Reproduced with permission of ref. 78. (b) Scheme of a twisted bilayer graphene fabricated as a 4-probe Hall bar device (top). 4-probe resistance $R_{xx}$ measured as a function of the temperature and carrier density for two different twist angles, 1.05° and 1.16°, very close to the 'magic angle = 1.1°'. Both graphs showed superconducting behaviour (bottom row) under the two angles, with variations in the superconducting domes. Reproduced with permission of ref. 90. (c) Scheme of the device, where the transition metal dichalcogenide monolayers have a twist angle close to zero (top). Real-space distribution of exciton centre-of-mass wavefunction. These maps show the apparition of three different moiré minibands (states I-III) in the strong-coupling regime (middle and bottom row). With permission of ref. 97.

Although it was predicted theoretically in 2011 [100], the experimental observation of unconventional superconductivity in two graphene layers twisted by the magic angle of about 1.1° in 2018, by Jarillo-Herrero's group, has shaken up the physics community. Figure 4(b) shows the 4-probe resistance $R_{xx}$ measured as a function of the carrier density and the temperature for two different devices. Both devices with the magic angles of 1.16° and 1.05°, respectively, display two superconducting domes, overlapped in the second case [101]. This work has boosted the emergence of the twistronics field [102,103], where all the rich electronic properties and range of magical angles in graphene twisted double-layers are being investigated [104–106]. On the other hand, the possibility to extend the study to other van der Waals layers twisted at small angles in order to produce novel collective excitation behaviours was soon realized. Different research groups have observed moiré interlayer excitons in $MoSe_2/MoS_2$ [107], $WSe_2/WS_2$ [108] and $MoSe_2/WSe_2$ [109] moiré superlattices. Jin *et al.* fabricated a fully h-BN encapsulated $WSe_2/WS_2$ heterostructure, with a measured twist angle of $0.5 \pm 0.3°$ (Figure 4(c), top). At the strong-coupling regime of a large moiré potential, three moiré exciton minibands appeared, represented in Figure 4(c), since under this regime the moiré potential is stronger than the exciton kinetic energy [108].

The influence of the interlayer interface in the physical properties is common in all type of two-dimensional van der Waals superlattices. A scanning tunnelling microscopy tip was employed to apply pressure on a graphene/h-BN, controlling the interlayer distance between the layers. It was demonstrated that the degree of commensurate stacking and the in-plane strain of graphene can be controlled as a function of the interlayer distance, tuning the electronic properties [110]. A laser induced shocking wave was used to eliminate wrinkles, bubbles and residues in the interlayer spaces of a graphene/BN/graphene moiré superlattices. This produced a reduction of the interlayer distance between the graphene layers and the boron nitride and an enhancement in the interlayer electron coupling. As a consequence, an opening of a small bandgap was observed [111]. It has been predicted theoretically that controlling the interlayer hybridization, a mosaic pattern with local topological insulator areas and normal insulator areas in moiré superlattices formed by two massive Dirac materials [112].

The results explained in this section show how an in-plane periodic potential can be introduced by stacking materials with a certain lattice mismatch or by twisting two 2D material layers under small angles. This method constitutes a novel approach to create superlattices with additional minibands, which display unconventional magnetic, superconducting, electronic and optical states.

### 2.4 Strain-engineered superlattices

Strain engineering is one of the most powerful strategies to modify the electronic and optical properties of 2D materials [113–116]. Strain is usually introduced unintentionally in many growth, mechanical stacking and device fabrication steps and its influence on the properties of the final superlattice should be addressed. But controlling the strain constitutes another route to tune the properties in superlattices and even to build up novel systems where periodic variation of strain leads to the fabrication of a superlattices.

In section 2.1, we described the interfacial phase-change memory materials based in $Sb_2Te_3$-GeTe vertical vdW superlattices. Zhou *et al.* improved the switching speed and lowered the energy consumption of their device by optimizing the synthesis of $Sb_2Te_1$-GeTe superlattices. The GeTe layers were grown on the $Sb_2Te_1$ layers subjected to a biaxial strain [117]. Another group tuned the level of induced biaxial strain on the GeTe layers of a $Sb_2Te_3$-GeTe 2D superlattice by varying the thickness of the $Sb_2Te_3$ layer [118]. Almost at the same time, three different groups [119–121] correlated the unexpected opening of gaps in graphene/h-BN moiré superlattices with the existence of corrugations or in-plane strain on the graphene and proposed theoretical models to explain this phenomena. San-Jose *et al.* [120] claimed that a displacement and an in-plane strain are produced in graphene due to the adhesion forces between the graphene and h-BN layers. They used known elastic constants for graphene and first-principles simulations to characterize the level of distorsion as a function of the twist angle. The strain/distorsion level represented as the local expansion (Tr(u)/2 in figure 5(a), top) was calculated for angles of 0°, 1.5° and 4° (Figure 5(a), top). They calculated the generated pseudomagnetic field for the same twist angles (Figure 5(a), bottom). A theoretical study showed that the strain along the three principal directions of a graphene layer can induced strong gauge fields that behaves as pseudo-magnetic fields, observing a new quantum Hall effect [122]. Lateral superlattices of alternating graphene/h-BN layers were studied theoretically to investigate the mismatch strain-relief mechanisms at the interface, which happened by misfit dislocations or by the formation of a periodic rippling as an out-of-plane relaxation [123]. The motivation behind this work was to point out the importance of controlling the experimental conditions to avoid the mismatch strain and obtaining coherent graphene/h-BN interfaces. The strain induced by the gold contacts on a graphene/h-BN moiré superlattice and its influence in the electrical properties has been studied as well[124].

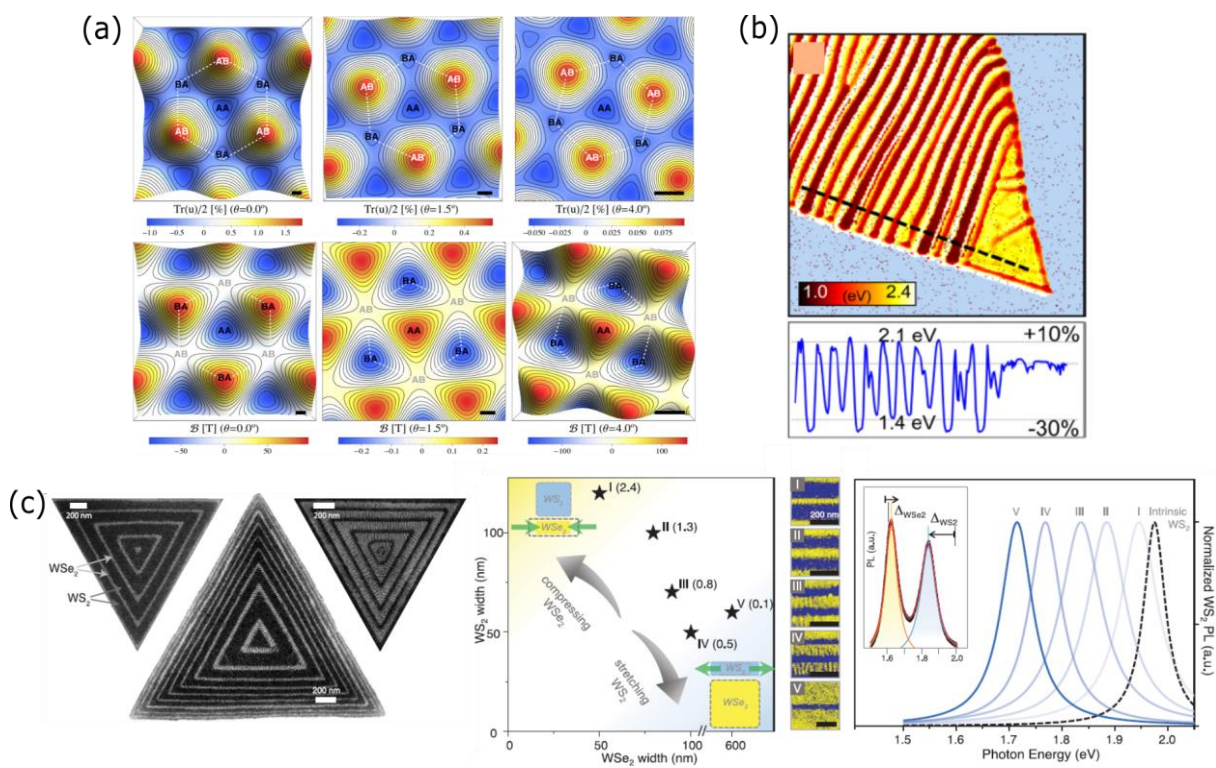

Figure 5 (a) Relative local expansion Tr$u$/2 simulated in real space for twist angles of 0º, 1.5º and 4º. These mappings represent the level of straining as a function of the rotation angle (top). Pseudomagnetic field distribution simulated in real space for the same twist angles of 0º, 1.5º and 4º. These mappings represent the fields that originate as a result of the level of straining (bottom). Reproduced with permission of ref. 109. (b) Iso-absorption map representing the energy at which the squared absorption coefficient is equivalent to 7.5x10$^{-11}$ cm$^{-2}$ at each point (top). A cross section along the dashed black line is shown in the bottom. The images show the existence of a spatially variable absorption spectra following the periodic compressive and tensile strain regions in the BP ripples. Reproduced with permission of ref. 114. (c) SEM images of three different WS$_2$/WSe$_2$ superlattices fabricated from coherent epitaxial growth (left). Graph with the aspect ratios (values in parenthesis) of the WS$_2$/WSe$_2$ widths at which the I to V superlattices were obtained and false-colour SEM images of the I-V superlattices (middle). Normalized PL spectra of the I-V superlattice together with the peak corresponding to the intrinsic WS$_2$ to show the shift in wavelengths (right). The inset spectrum of a typical WS$_2$/WSe$_2$ superlattice is represented to illustrate the relative WS$_2$ peak red-shift and the relative WSe$_2$ peak blue-shift. Reproduced with permission of ref. 115.

Tuning the bandgap of semiconducting 2D materials is one of the main focus of strain engineering, relevant for optoelectronic applications. The formation of periodic ripples on a black phosphorus thin layer gave rise to an absorption spectra with periodic absorption edge shifts of +10% and -30%, corresponding to the regions under tensile and compressive strains [125]. An iso-absorption map shown in Figure 5(b) illustrates the spatially varying optical properties induced by the periodic ripple of the black phosphorus flake. Coherent, full-matching lattice constants in-plane $WS_2/WSe_2$ superlattices were fabricated by metal-organic chemical vapor deposition. A SEM image of three different superlattices are shown in figure 5(c) (left). The lattice coherence has been achieved by tensile (compressive) strain in the $WS_2$ ($WSe_2$) components. The ratio of their widths allows to control the degree of strain, producing different type of superlattices (labelled as I to V within the panel) (figure 5(c), middle). Each type of superlattice (I to V) presents a red- (blue-) shifted peak in the photoluminiscence spectra with respect to the intrinsic $WS_2$ ($WSe_2$). The spectra is shown in figure 5(c) (right) and states the potential to use this approach for tunable optoelectronic devices [126]. The misfit dislocations formed at the interface of a lateral $WSe_2/WS_2$ heterostructure as the result of their lattice mismatch were used as templates to grow $WS_2$ sub-2 nm quantum wells on a $WSe_2$ matrix superlattice by chemical vapour deposition. The strain field surrounding the misfit dislocations was found to induce dislocation climb by the preferential growth of the quantum wells[127].

Rigorously, strain is the mechanism that is common in almost all the types of superlattices described in this *Feature article* through the lattice mismatch or the stress induced on the 2D material layers when they are stacked together or placed in a patterned topography. For this reason, a deep study and understanding of its influence in the fabrication of superlattices in particular, and any other device in general, will be fundamental for the advance of the 2D materials field.

**2.5 Lithographic-engineered superlattices**

This section includes those works were the periodicity of the superlattice was created by either transferring a 2D material onto a substrate with an ordered array of structures that served as templates to tailor its electronic and optical properties or patterning selectively a 2D material layer by a lithographic method.

The top panel of figure 6(a) shows a schematic of a $MoS_2$ monolayer transferred onto an array of periodic $SiO_2$ nanocones, which are assembled by nanosphere lithography. At the apex of the nanocones, the more stretched areas of the $MoS_2$ present a high tensile strain, which results in a periodic decrease in its optical bandgap. Therefore, a large area artificial-atom crystal is produced. Raman and photoluminescence measurements showed that crystal presented a large exciton binding energy. Figure 6(a) (bottom left) shows a scanning PL map of the artificial crystal. Under illumination with wavelength larger than the bandgap of $MoS_2$, excitons are created in $MoS_2$. These excitons drift towards the nanocone tips and then, are emitted with a larger wavelength. A scheme of the process is shown in figure 6(a) (bottom right)[128]. Another group employed an alternative approach to fabricate artificial superlattices based on a $MoS_2$ monolayer. They generated a periodical array of quantum dots inducing local phase change from 2H (semiconducting) to 1T (metallic) by focused electron beam exposure. By controlling the size and the pitch of the quantum dots, the tuning of the bandgap from 1.81 to 1.42 eV was demonstrated [129]. An artificial superlattice was achieved by transferring a graphene layer onto a self-assembled array of silicon oxide nanoparticles. The quasi-periodic strain produced on the graphene gave rise to minibands with broad density of states. The authors claimed that this result constituted a cheap and simple process to fabricate superlattices for optical modulators or infrared sensors [130]. In another work, a graphene layer was transferred onto an array of gold nanoislands. A thermal annealing creates an hybrid graphene/gold superstructure where the stretching of the graphene and the phase change in gold gave rise to anormal large moiré periodicities[131].

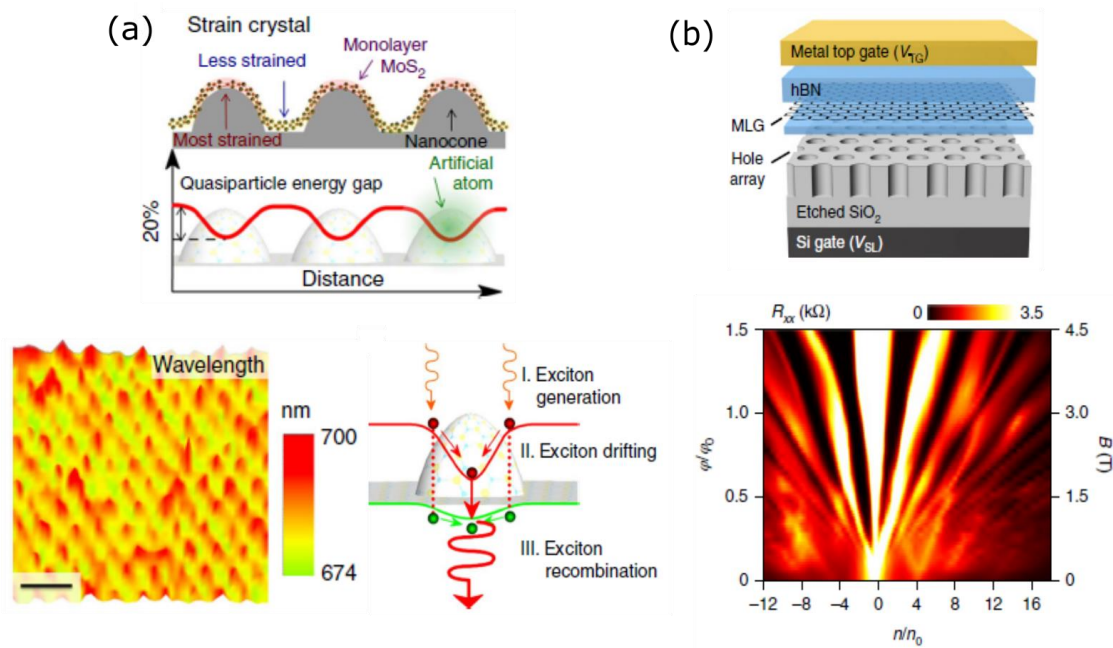

Figure 6. (a) Scheme of the artificial-atoms crystal by transferring of $MoS_2$ monolayer on an array of periodic $SiO_2$ nanocones. The energy bandgap reduction is inverse to the strain level (top). Scanning PL spectroscopy mapping of the artificial crystal (bottom left). Scheme of the process that the excitons undergo once they are induced on the artificial crystal under light exposure (bottom right) Reproduced with permission of ref. 117. (b) Scheme of a final device consisting in a full encapsulated graphene transferred on a patterned dielectric superlattice (top). Device $R_{xx}$ resistance measured as a function of normalized density and magnetic field. A Hofstadter butterfly behaviour is achieved. Conditions: $V_{superlattice}$ = 50V, T = 250 mK, application of a triangular periodic potential (bottom). Reproduced with permission of ref. 122.

The previous section showed results where moiré patterning was employed to tune the bandgap structure of graphene. A second route to achieve this is by nanostructuring the material by lithographic methods. In a recent work, an array of triangular shape holes with a pitch of 35 nm and minimum feature sizes of 12-15 nm was defined on a h-BN/graphene/h-BN heterostructure by electron beam lithography. The small size and the high density of the holes were chosen to manipulate the electronic and magnetic quantum transport behaviour [132] A third route to engineer the electronic properties of graphene is by applying external electric fields rather than modify the material itself. A fully encapsulated h-BN/graphene/h-BN heterostructure was transferred onto a 300 nm $SiO_2$ substrate with an array of periodic nanoholes defined by electron beam lithography. A scheme of the device is shown in figure 6(b), top. This patterned dielectric layer was the component that induced a superlattice structure on graphene, by introducing periodic potentials with triangular or square unit cells. As a result, Dirac cones and Hofstadter butterfly behaviours were replicated from non-moiré generated superlattices. Figure 6(b) (bottom) shows a Hofstadter's butterfly spectrum achieved at 4.5 T, while the observation of the Hofstadter spectrum from moiré superlattices requires magnetic fields higher than 20 T up to now [133].

Lithography is ubiquitous in the fabrication of devices. However, in the case of 2D material based superlattices, the results from the examples shown in this section can be obtained by using one of the methods described in the previous sections. This poses the question, whether the lithographic-engineered approach is well-suited to produce optimal van der Waals based superlattices in comparison with the other strategies, since the mixed dimensionality of a 2D material flake and the 3D pattern substrate can introduce an additional complexity at the interlayer interaction and expensive and complex techniques such as electron and focused ion beam are required to define high-resolution periodic potentials on a 2D material surface.

# 3 Conclusion and Outlook

2D materials and their van der Waals interlayer interaction display physical and chemical properties unobserved in their bulk counterparts. This unique trait has opened the way to produce devices with novel architectures, among them, the fabrication of van der Waals 2D materials superlattices. Our aim in the present *Feature article* was to underscore the huge potential and relevance of this type of structures, which comprise a very young field yet. The first prototypes have shown improved thermoelectric, optoelectronic, phase-change memory and energy storage performances, compared to the devices fabricated with established methods. These achievements constitute an invitation to explore the two-dimensional based superlattices for solving environmental and energy consumption related issues. Particularly, they could have a great impact in the thermoelectrical and Li-/Na- storage applications. At the fundamental level, the Moiré superlattices have displayed novel electronic, excitonic, superconductive, magnetic states. The field of twistronics have just started and a plethora of new physical phenomena is expected in the future. Strain is present in almost all the types of superlattices, both introduced unintentionally or as a variable to modulate their properties. Therefore, a rigorous study of the strain influence/engineering in the final properties of the structures will be fundamental for the controlled fabrication of high-quality superlattices. The superlattices based on van der Waals 2D materials have shown a huge versatility. As an example, the bandgap can be modulated by the composition of the superlattice unit cell, controlling the interlayer interaction, intercalating ions or molecules at the interlayer, moiré patterning, strain engineering, combining 2D materials with conventional lithographic approaches. This illustrates the potential to design and devise superlattices. For the same reason, many research groups from different chemical and physical backgrounds can access to their fabrication. Finally, the improvement and expansion of the robotic assembly approach will be crucial for the viability of 2D material based superlattices both at the academic and industrial level. At the academic level, the automation would allow the researchers saving time and increasing the success yield in the fabrication of complex superlattices, in order to focus in the study and engineering of their properties. The robotic assembly has room to introduce more variables such as the twist angle, the doping level or the cleaning of the flakes, which will result in the production of high-quality superlattices. At the industrial level, the remaining challenge is the implementation of the large-scale manufacturing of these superlattices.

**Acknowledgements**

This project has received funding from the European Research Council (ERC) under the European Union's Horizon 2020 research and innovation programme (grant agreement n° 755655, ERC-StG 2017 project 2D-TOPSENSE). R.F. acknowledges the support from the Spanish Ministry of Economy, Industry and Competitiveness through a Juan de la Cierva-formación fellowship 2017 FJCI-2017-32919